\documentclass[final,3p,times]{elsarticle}

\usepackage{amssymb}
\usepackage{amsmath}
\usepackage{amsfonts}
\usepackage{gensymb}
\usepackage{siunitx} 

\usepackage{float}
\usepackage{amsmath}
\usepackage{hyperref}
\newcommand{\etal}{et al. } 

\usepackage[utf8]{inputenc}
\usepackage{newunicodechar}
\newunicodechar{θ}{\ensuremath{\theta}}
\usepackage{newunicodechar}
\newunicodechar{𝑎}{a}

\journal{Computational Material Scinece}

\begin{document}

\begin{frontmatter}



\title{Torsional Behavior of Carbon-Doped Ferrous Nanowires: Atomic-Scale Insights from MD Simulations}

\author{Charith L. Hirimuthugodage} 

\affiliation{organization={Department of Applied Computing, Faculty of Computing and Technology, University of Kelaniya},
	city={Kelaniya},
	country={Sri Lanka}}

\author{Laalitha S.I. Liyanage \corref{cor1}} 
\ead{laalitha@kln.ac.lk} 
\cortext[cor1]{Corresponding author} 

\affiliation{organization={Department of Applied Computing, Faculty of Computing and Technology, University of Kelaniya},
	city={Kelaniya},
	country={Sri Lanka}}

\author{K. G. S. H. Gunawardana}
\affiliation{organization={Department of Engineering Technology, Faculty of Technology, University of Ruhuna},
	country={Sri Lanka}}

\begin{abstract}

This study investigates the torsional mechanical properties of pristine iron (Fe) and carbon-doped iron (FeC) nanowires with [001] orientation through molecular dynamics simulations utilizing the Modified Embedded Atom Method (MEAM) potential developed by Liyanage et al. for accurately modeling Fe-C interactions in body-centered cubic structures. Systematic analysis across carbon concentrations (0 - 10\%), temperatures (1 - 900 K), and cross-sectional dimensions 10a, 13a, 15a, ( where a = 2.81 Å represents the lattice constant ) within the LAMMPS environment reveals that increasing carbon content weakens grain boundaries, reducing the maximum shear stress required to reach the critical torsional angle, while higher temperatures promote phase transitions from elastic to plastic deformation due to enhanced atomic vibrations, and larger cross-sections exhibit higher shear stress resistance attributed to strengthening effects from the outer atomic layers. By elucidating these interrelationships between carbon content, temperature, and dimensional factors, this work provides fundamental insights into the mechanical behavior of FeC nanowires under torsional loading conditions, offering valuable guidance for their potential applications in nanoelectromechanical systems, nanorobotic actuators, and advanced structural materials where precise control of mechanical properties is essential.

\end{abstract}

\begin{graphicalabstract}
\end{graphicalabstract}

\begin{highlights}
\item \textbf{Atomic-Scale Insights into Carbon Doping Effects:} This study reveals that carbon doping weakens the torsional load-bearing capacity of ferrous nanowires, with higher carbon concentrations diminishing the material's shear strength and disrupting its elastic-plastic behavior. These findings provide crucial atomic-scale understanding of impurity-driven mechanical property degradation.

\item \textbf{Comprehensive Impact of Temperature and Size Effects:} The research demonstrates that increasing temperature reduces torsional strength due to enhanced lattice instability, while larger cross-sectional sizes improve shear stress tolerance through delayed fracture propagation. This dual analysis offers predictive insights for designing temperature-resilient and size-optimized nanostructures.

\end{highlights}


\end{frontmatter}



One-dimensional metallic nanostructures  have received extensive attention due to their unique electrical, optical, and mechanical properties \cite{khalid2020advanced} , and have been applied to various micro/nano devices. Among these nanowires, carbon-doped iron nanowires have emerged as critical components with diverse applications due to their enhanced performance characteristics \cite{kehoe1970role}. In these devices, nanostructures prepared as structural components are subjected to miscellaneous loadings inevitably. It is fundamental to study and comprehend the mechanical properties and deformation behaviors of these nanostructures, which provides a good reference for robust device design.

Typical loading measurements like torsion on nanowires are hard to carry out in the laboratory, while molecular dynamics (MD) simulations provide a feasible method to implement the examinations and observe the details during loadings. By modeling the response of nanostructures to torsional stress, these computational studies provide valuable insights into their behavior. Extensive molecular dynamics (MD) investigations have been conducted on the mechanical properties of iron-based nanowires under tensile loading conditions. Nadeesha~\emph{et al.}~\cite{nadeesha2021prediction} systematically examined the tensile mechanical properties of pure iron and iron-carbon nanowires, demonstrating that the mechanical behavior exhibits strong dependence on carbon concentration and temperature. In complementary work, Sainath~\emph{et al.}~\cite{sainath2017atomistic} investigated the tensile deformation and fracture mechanisms of [111] body-centered cubic (BCC) Fe nanowires, revealing that at low temperatures (10--375~K), nanowires undergo yielding through sharp crack formation and exhibit predominantly brittle failure modes.

Comparatively torsional deformation studies of Fe-C nanowires remain absent from the literature. The available torsional investigations have focused primarily on other metallic nanowire systems. Yang~\emph{et al.}~\cite{yang2021molecular, yang2022comparisons} conducted comprehensive studies on the torsional behavior of copper-aluminum core-shell nanowires and pure copper nanowires, identifying that dislocation ring structures at the interface significantly enhance the stability of core-shell nanowires despite lattice mismatch. Their findings further revealed that while torsion rates do not substantially influence elastic properties, a modest strengthening effect is observed at elevated torsion rates. Additionally, Yang~\emph{et al.}~\cite{yang2018molecular} investigated the electro-plastic behavior of copper nanostructures under torsion, establishing that the strength of single-crystal copper nanowires increases with extended length, reduced torsion rate, and decreased operating temperature. Further studies include investigations on copper nanowires by Gao~\emph{et al.}~\cite{gao2010investigation}, aluminum nanowires by Sung~\emph{et al.}~\cite{sung2012effects}, and six-fold twinned $\alpha$-ferrite nanowires by Li~\emph{et al.}~\cite{li2017large}. While the aforementioned studies provide valuable insights into various mechanical properties of iron-based and other nanowires, the torsional behavior of carbon-doped iron nanowires remains largely unexplored, highlighting a significant gap in the literature.

In this work, we investigate the torsional properties of Fe-C nanowires. This study employs molecular dynamics simulations to examine the effects of carbon doping concentration, temperature, and cross-sectional size on the torsional behavior of these nanostructures. By analyzing the atomic-scale mechanisms governing the torsional strength and failure of these nanostructures under varying conditions, we aim to provide insights that can aid in designing stronger and more efficient nanodevices for nanoelectromechanical systems (NEMS), detectors, actuators, and energy storage applications.

\section*{Methodology}

Large scale atomic/molecular massively parallel simulator (LAMMPS) \cite{plimpton2012computational} is employed to perform the MD simulations. The modified embedded atom method (MEAM) potential for Fe-C systems, developed by Liyanage \etal \cite{liyanage2014structural}, is chosen to describe the interaction among atoms, which has been widely used in Fe-C nanowire studies \cite{nadeesha2021prediction} and accurately describes the body-centered cubic (BCC) iron structure, Fe-C systems such as cementite (Fe3C), L12 and B1 structures, as well as C interstitials in Fe BCC lattice.

The whole model, placed in the center of a simulation box, is constructed with circular cross-section nanowires with ferrous atoms arranged in a perfect BCC lattice. The lattice orientations of [100], [010], and [001] are in the direction of the X, Y, Z axes of the Cartesian coordinate. Both ends of the model (colored in yellow) are the load regions, and the middle section (colored in red) is the operating part Figure \ref{fig 2}. For the initial configuration with a radius of 10a and a length of 100a (0\% carbon doping), the system consists of approximately 15,900 atoms. The load regions are set as rigid bodies, while the atoms of the operating part are allowed to move freely without any stress constraints to ensure the nanowire deforms naturally during loading. Carbon atoms are randomly distributed in the iron lattice at concentrations with different concentrations. Shrink-wrapped boundary conditions are applied to the three axial directions, and the time step is 1 femtosecond.

The simulation process constitutes three sequential stages: energy minimization (I), thermal equilibration (II), and torsion loading (III). Initially, the nanowire undergoes energy minimization using the conjugate gradient method \cite{harden2008convergence} to eliminate any unfavorable atomic configurations. Then the atoms are assigned initial velocities randomly according to the Maxwell-Boltzmann distribution and equilibrated for 100,000 steps (100 ps) using the NVT canonical ensemble with the Nose-Hoover thermostat \cite{miyazaki1996calculation} to attain equilibrium at the specified temperature. Subsequently, torsional loading is applied by rotating both end regions in opposite directions (the top layer revolves anticlockwise and the bottom layer clockwise when observed from the top) around the Z-axis at a constant angular velocity of 2 × 10$^{11}$ degrees/second. After each 1- degree rotation, the system is allowed to relax for 20 ps to ensure the nanowire attains a new stable state before further loading. This step-wise loading process continues until structural failure occurs, with the canonical ensemble (NVT) maintaining constant temperature throughout the entire process.

Simulations are performed across different temperatures (1 K, 300 K, 600 K, 900 K), cross-sectional diameters (10a, 13a, 15a, corresponding to approximately 15,900, 26,100, and 34,900 atoms), and carbon concentrations (0\%, 1\%, 5\%, 10\%) to systematically examine their influences on mechanical properties. During the simulation, atomic configurations are visualized using OVITO \cite{stukowski2009visualization} to observe deformation mechanisms and defect formation. The shear stress is calculated as the sum of the per-atom stress components calculated by the Virial expression implemented in LAMMPS, containing terms related to both kinetic energy from temperature vibration and potential energy from lattice deformation. The torsional angle (θ) is defined as the relative rotation angle between the two end regions, measured in degrees, with the critical torsional angle determined as the point at which maximum shear stress is reached before plastic deformation begins.

\begin{figure}[H] \centering \includegraphics[width=0.5\textwidth]{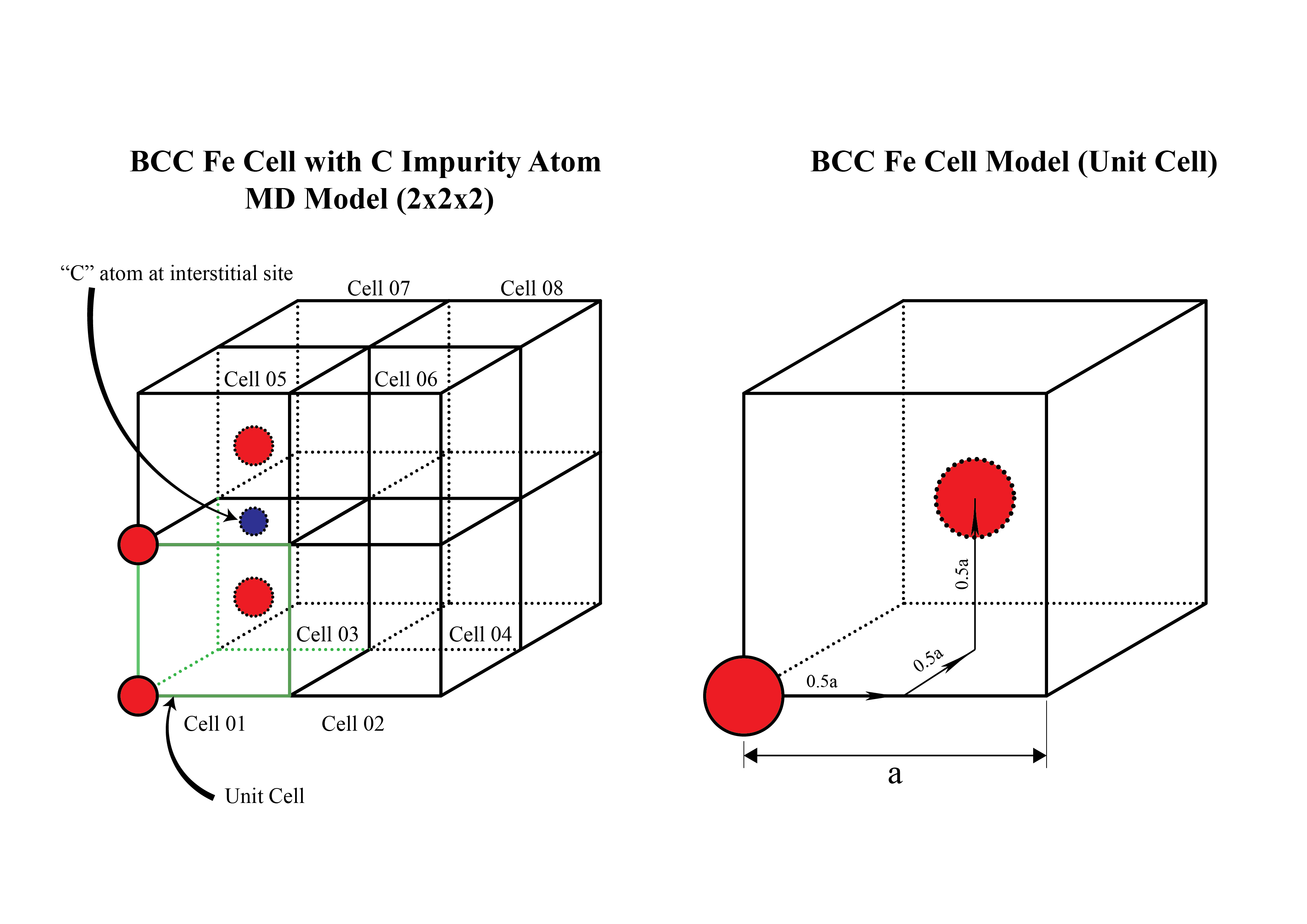} \caption{BCC Fe unit cell with carbon impurity in interstitial position.} \label{fig 1} \end{figure}

\begin{figure}[H] \centering \includegraphics[width=0.15\textwidth]{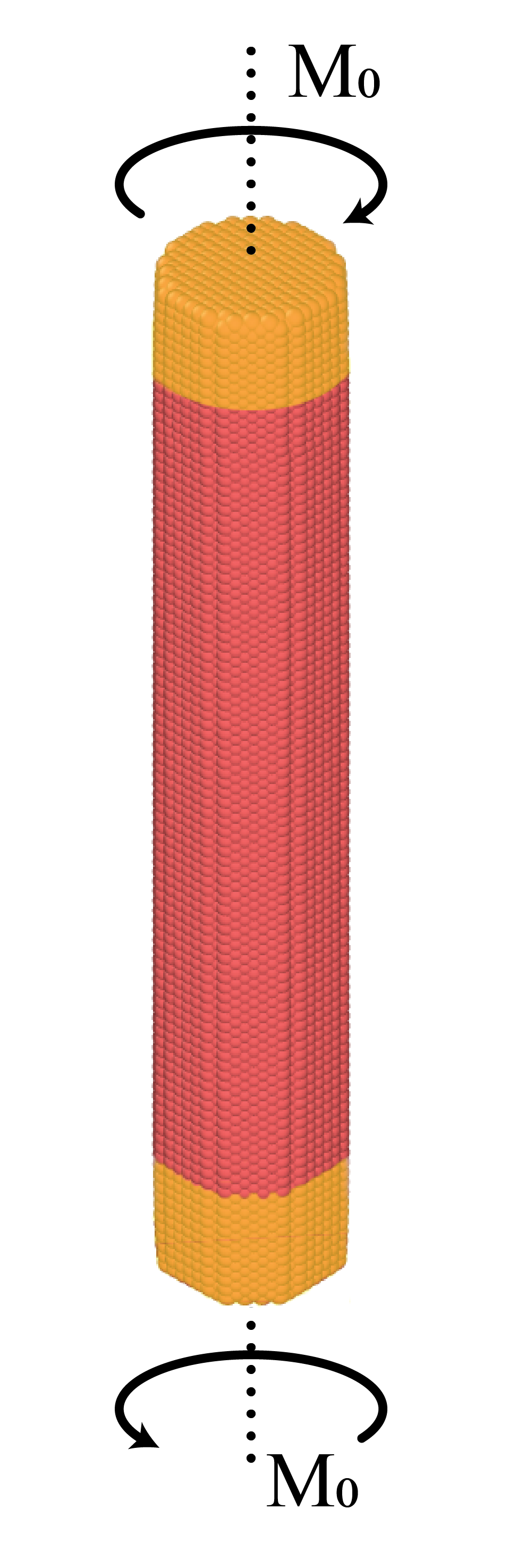} \caption{Schematic of the Fe nanowire model. Yellow regions indicate torsional motion, and red regions deform naturally.} \label{fig 2} \end{figure}
\newpage

\section*{Simulation Results and Discussion}

\textbf{Effect of Carbon Concentration on Torsion} \\

\begin{figure}[H]
	\centering
	\includegraphics[width=0.5\textwidth]{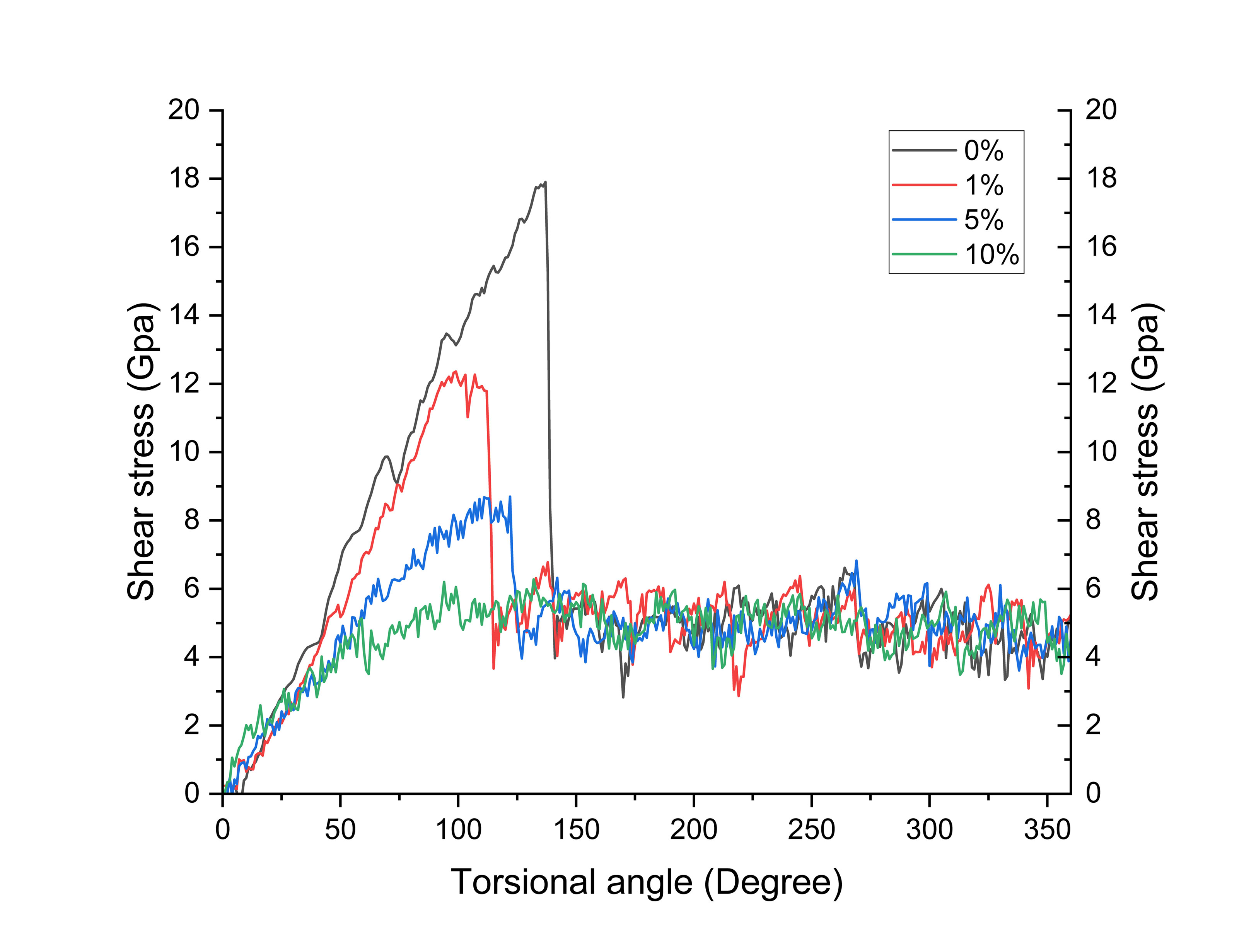}
	\caption{Shear stress vs. torsional angle for 10a nanowire at 1 K. Each curve corresponds to a specific carbon concentration (0\%, 1\%, 5\%, 10\%)}
	\label{FIG:3}
\end{figure}

\begin{figure}[H]
	\centering
	\includegraphics[width=0.5\textwidth]{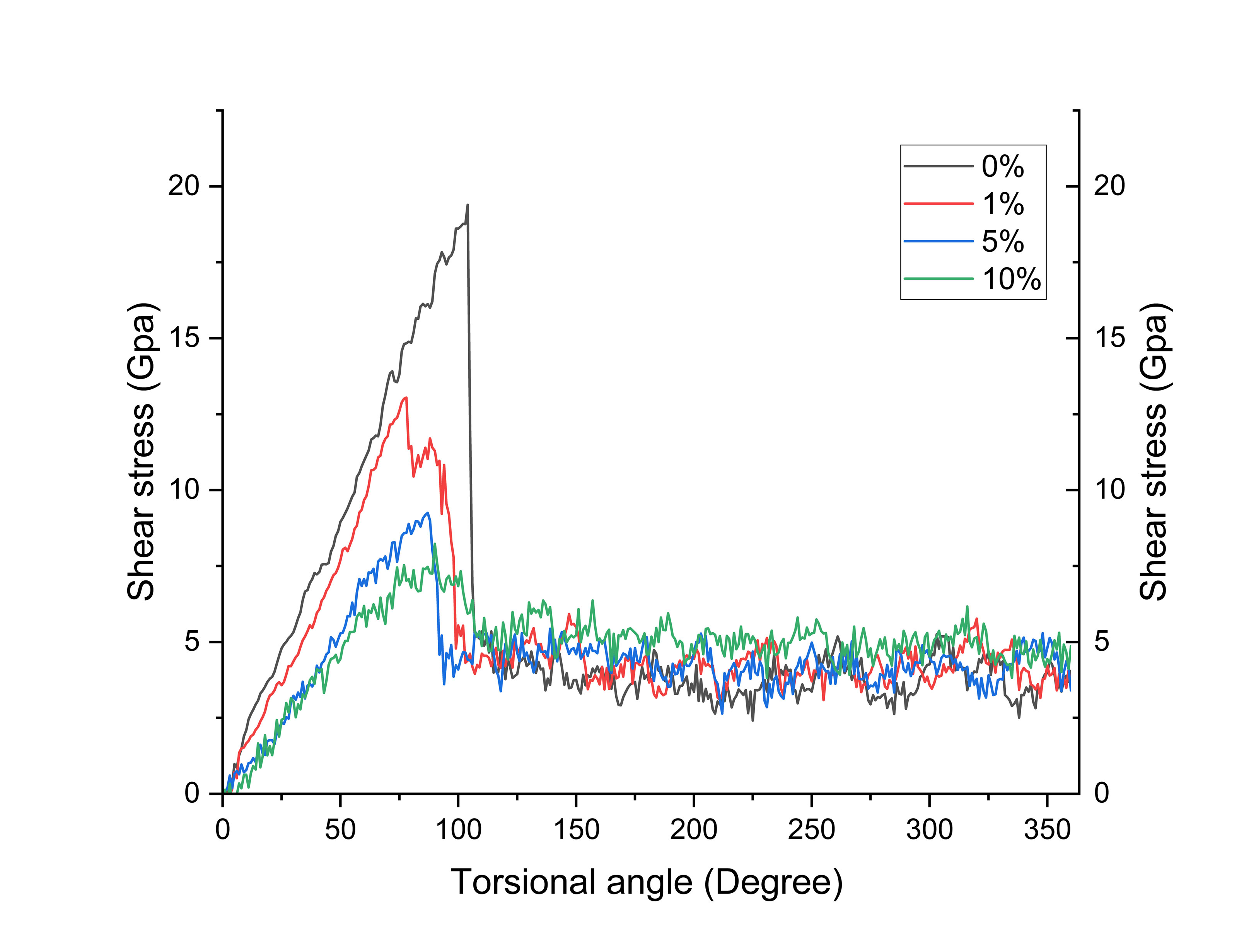}
	\caption{Shear stress vs. torsional angle for 13a nanowire at 1 K. Each curve corresponds to a specific carbon concentration (0\%, 1\%, 5\%, 10\%)}
	\label{FIG:4}
\end{figure}

\begin{figure}[H]
	\centering
	\includegraphics[width=0.5\textwidth]{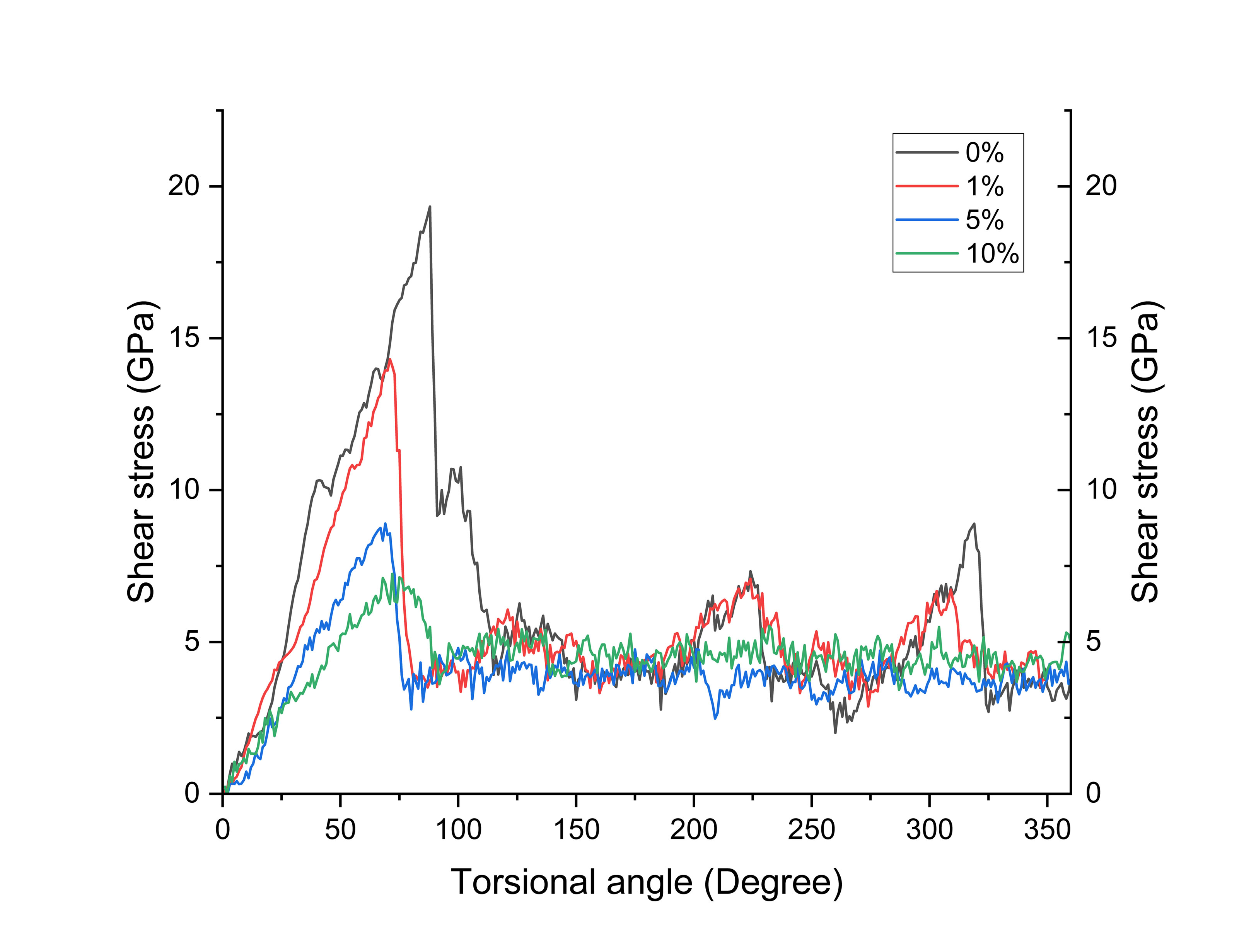}
	\caption{Shear stress vs. torsional angle for 15a nanowire at 1 K. Each curve corresponds to a specific carbon concentration (0\%, 1\%, 5\%, 10\%)}
	\label{FIG:5}
\end{figure}

To examine the effect of carbon concentration on the torsional behavior of Fe–C nanowires, simulations were performed on nanowires with carbon contents of 0\%, 1\%, 5\%, and 10\% under constant temperature conditions. Three cross-sectional sizes—10a, 13a, and 15a—were considered, each with a fixed length of 100a. The shear stress versus torsion angle curves for these nanowires are shown in Figure \ref{FIG:3} , \ref{FIG:4} and Figure \ref{FIG:5}, respectively. For the 10a cross-section nanowire (Figure \ref{FIG:3}), the results reveal that increasing carbon content leads to a progressive reduction in the critical torsion angle, which marks the transition from elastic to plastic deformation. The carbon-free nanowire exhibits a critical angle of approximately 137°, indicating a stable elastic regime prior to plastic flow. However, with the introduction of 1\% and 5\% carbon, the critical angle reduces, and beyond 5\%, at 10\% carbon concentration, the torsional response becomes unstable, making it difficult to clearly distinguish between elastic and plastic deformation regions.

 A similar trend is observed in the 13a nanowire (Figure \ref{FIG:4}), where higher carbon content results in a reduced critical angle and lower shear stress capacity. The carbon-free nanowire in this case also shows the highest mechanical resistance, while the 10\% carbon-doped nanowire shows significant instability under torsional load. In the case of the 15a nanowire (Figure \ref{FIG:5}), consistent behavior is observed, with carbon-free structures sustaining higher torsion angles and shear stresses, while increased carbon doping leads to a degradation of torsional performance. This reduction in mechanical stability is attributed to lattice distortions caused by carbon impurities, which act as stress concentrators and promote early dislocation nucleation. 
 
\begin{figure}[H]
	\centering
	\includegraphics[width=0.8\textwidth]{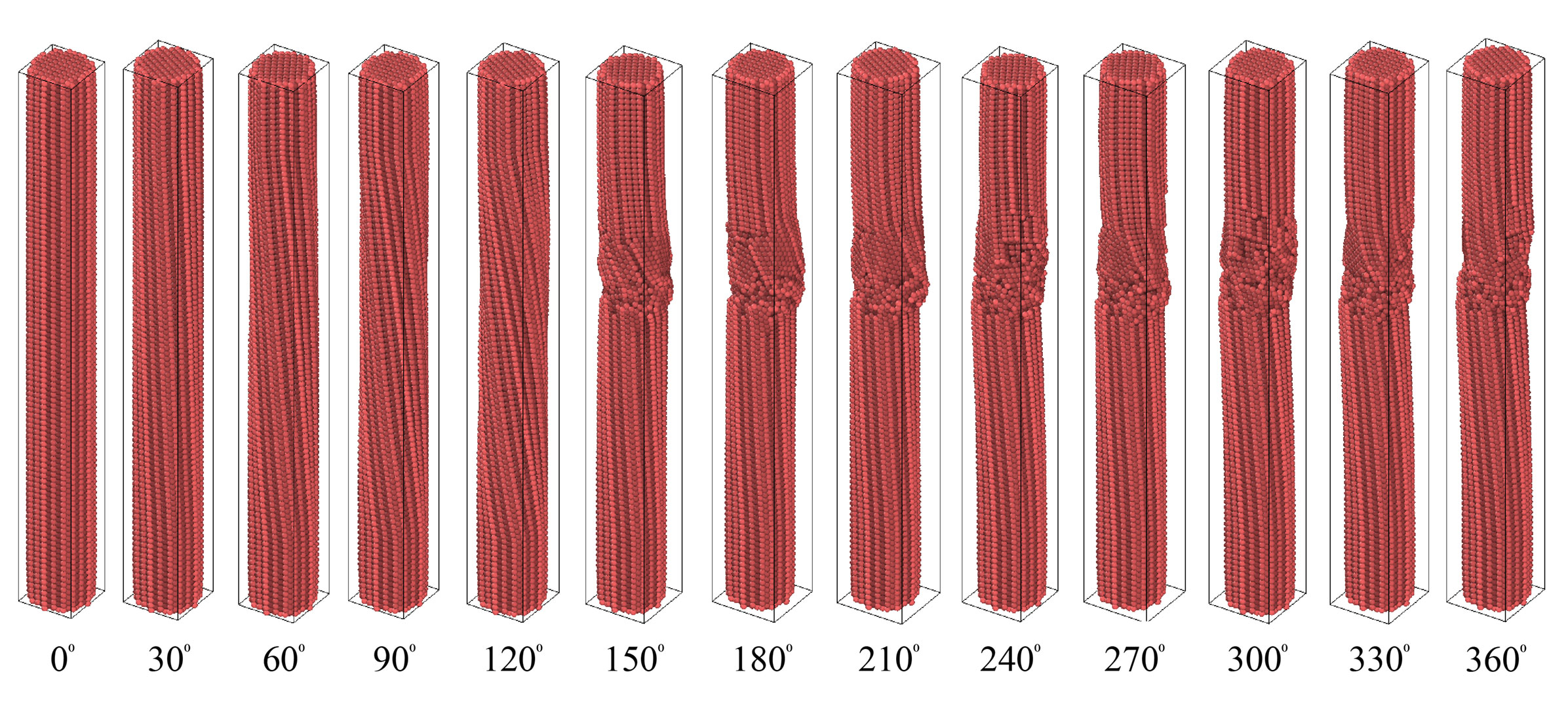}
	\caption{Dislocation structures of the nanowire with a radius of 10 at different torsional angles}
	\label{FIG:6}
\end{figure}

 In Figure \ref{FIG:6}, further illustrates the atomic-scale deformation during torsion for a 10a nanowire with 0\% carbon content at 1 K, analyzed using Common Neighbor Analysis (CNA). Red atoms represent the BCC lattice, and brown atoms indicate dislocations. The deformation initiates from the outer surface and propagates inward, with dislocations becoming increasingly prevalent as the torsion angle exceeds the critical threshold. In the elastic region, the atomic structure remains largely intact, but upon entering the plastic regime, dislocation density increases sharply, leading to significant structural rearrangement. The simulation results clearly show that the increase in carbon concentration reduces the nanowire’s ability to sustain elastic deformation and promotes early plasticity and structural instability. At high doping levels, especially above 5\%, the nanowires exhibit indistinct deformation stages, indicating that excessive carbon content undermines mechanical integrity during torsion.

\begin{figure}[H]
	\centering
	\includegraphics[width=0.5\textwidth]{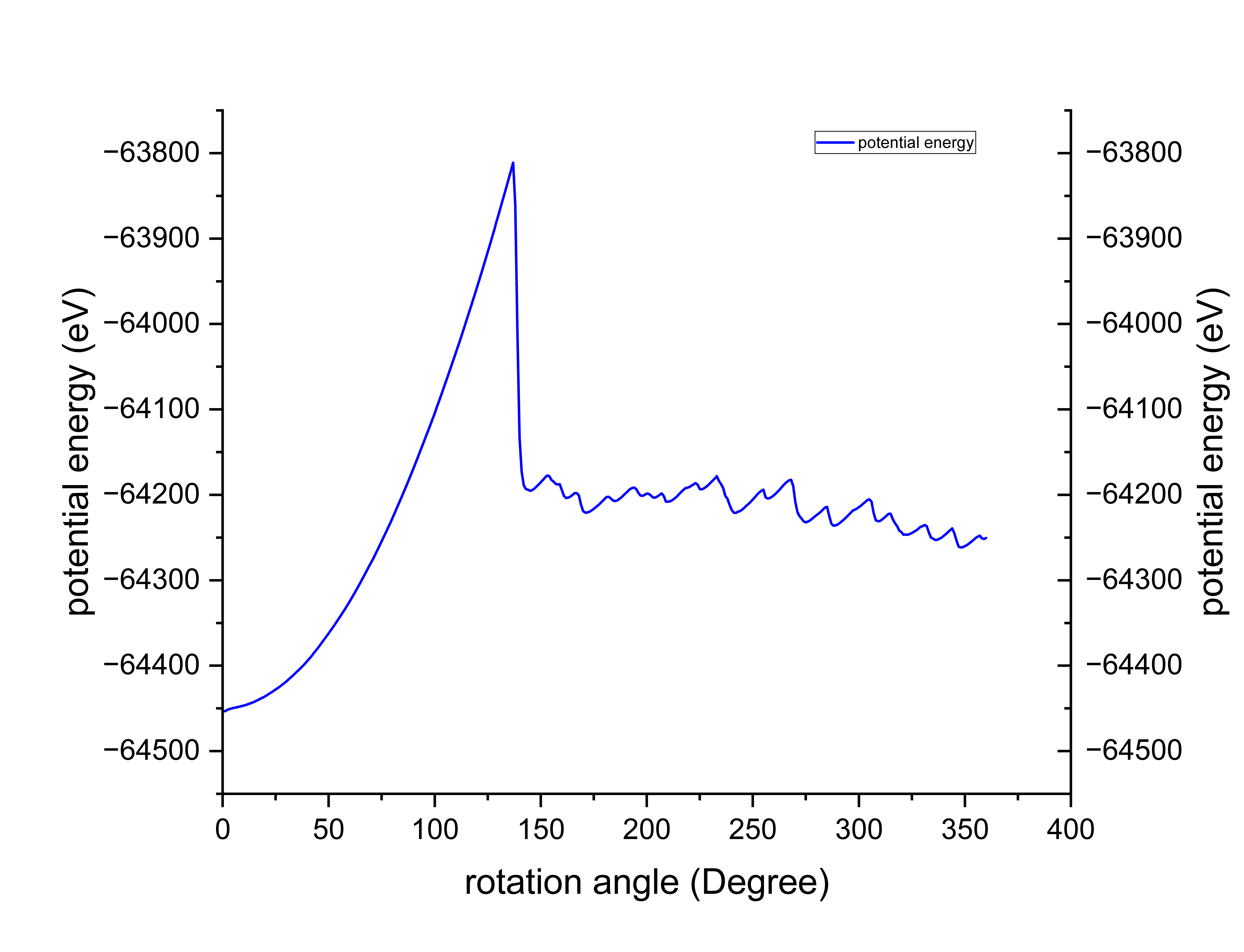}
	\caption{Potential Energy vs. Rotation Angle for 0 \% Carbon Content in a 10a Diameter Nanowire at 1 K}
	\label{FIG:7}
\end{figure}

\textbf{Effect of Temperature on Torsion}\\

The influence of temperature on the torsional response of Fe-C nanowires was analyzed using a nanowire model with a length of 100a ,a  circular cross-sectional diameter of 10a, and a torsional loading rate of 2 × 10$^{11}$ degrees/second. Simulations were conducted at 0 K, 300 K, 600 K, and 900 K. As the temperature increases, intensified atomic thermal motion leads to larger fluctuations in the potential energy, with noticeably higher fluctuation amplitudes observed above 300 K. During the elastic deformation stage, elevated temperatures reduce the amplitude of potential energy changes, indicating that less energy is required for deformation. This behavior corresponds to lower critical torsional angles at higher temperatures, suggesting a weakening of the mechanical resistance of the nanowire.
\\

\begin{figure}[]H
	\centering
	\includegraphics[width=0.5\textwidth]{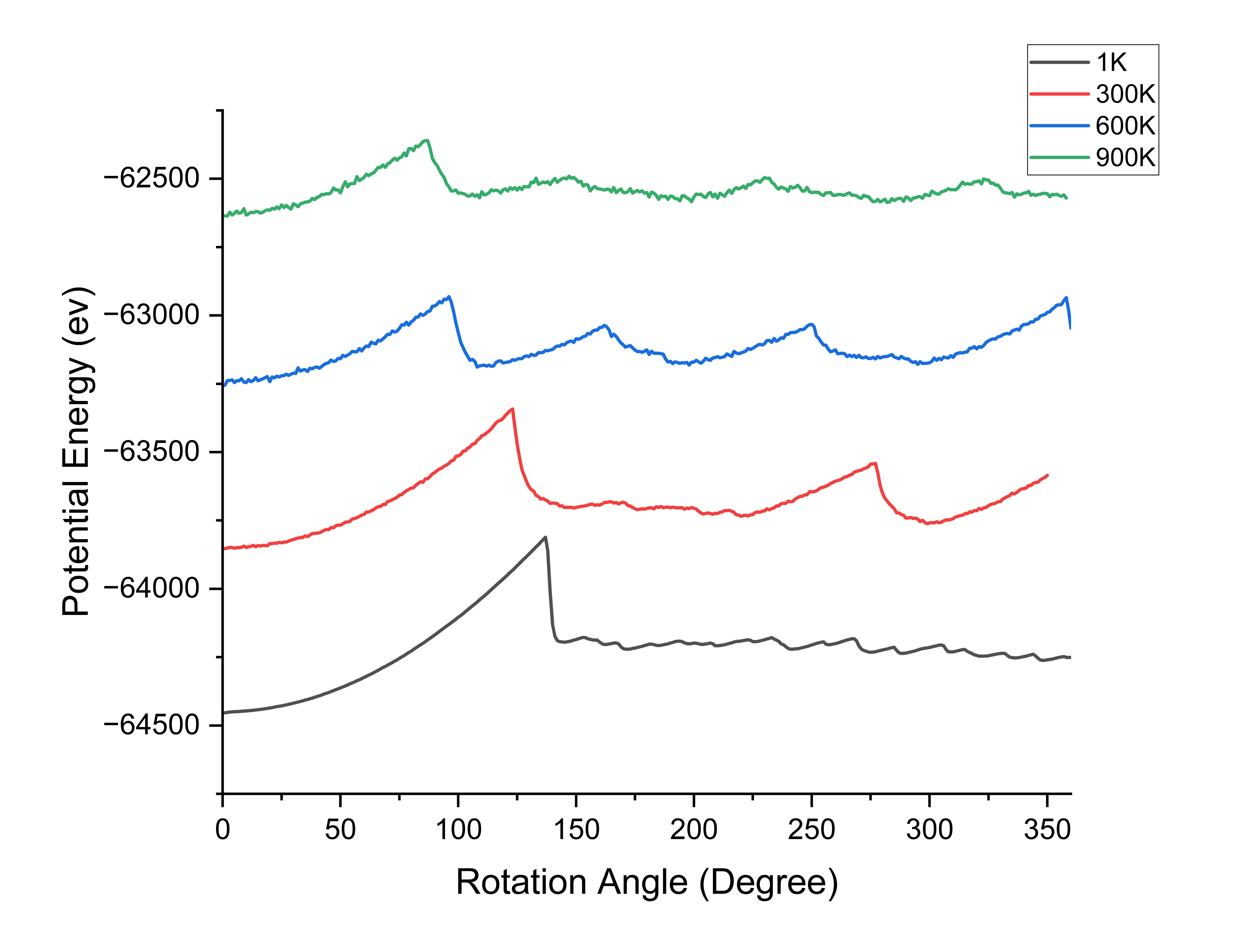}
	\caption{Potential energy vs. rotation angle for 0\% carbon with a 10 \(a\) diameter.}
	\label{tmp05}
\end{figure}

\begin{figure}[H]
	\centering
	\includegraphics[width=0.5\textwidth]{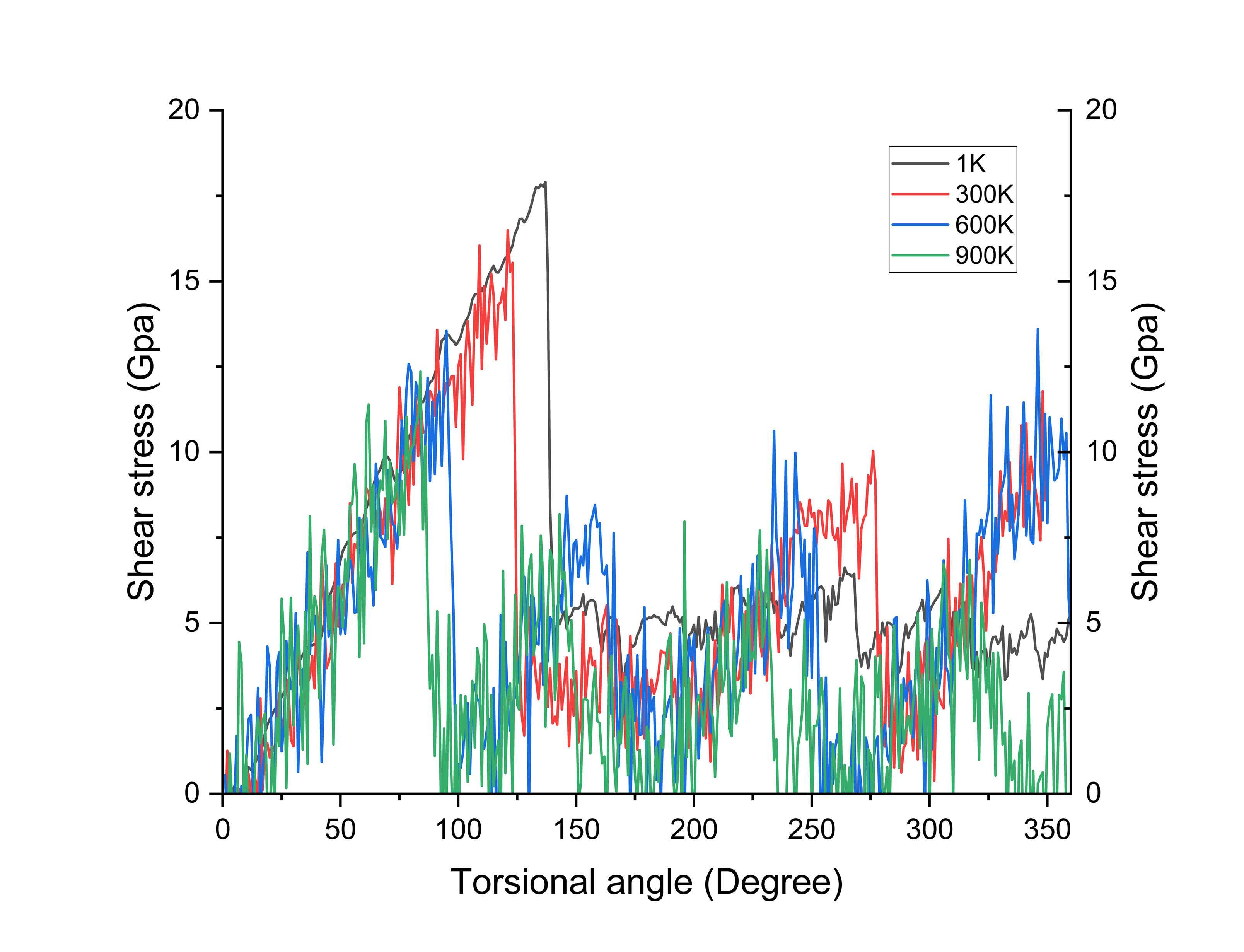}
	\caption{Shear stress vs. torsional angle for a 10\(a\) nanowire with 0\% carbon concentration. Each curve corresponds to a specific temperature (1 K, 300 K, 600 K, and 900 K}
	\label{tmp01}
\end{figure}

\begin{figure}[H]
	\centering
	\includegraphics[width=0.5\textwidth]{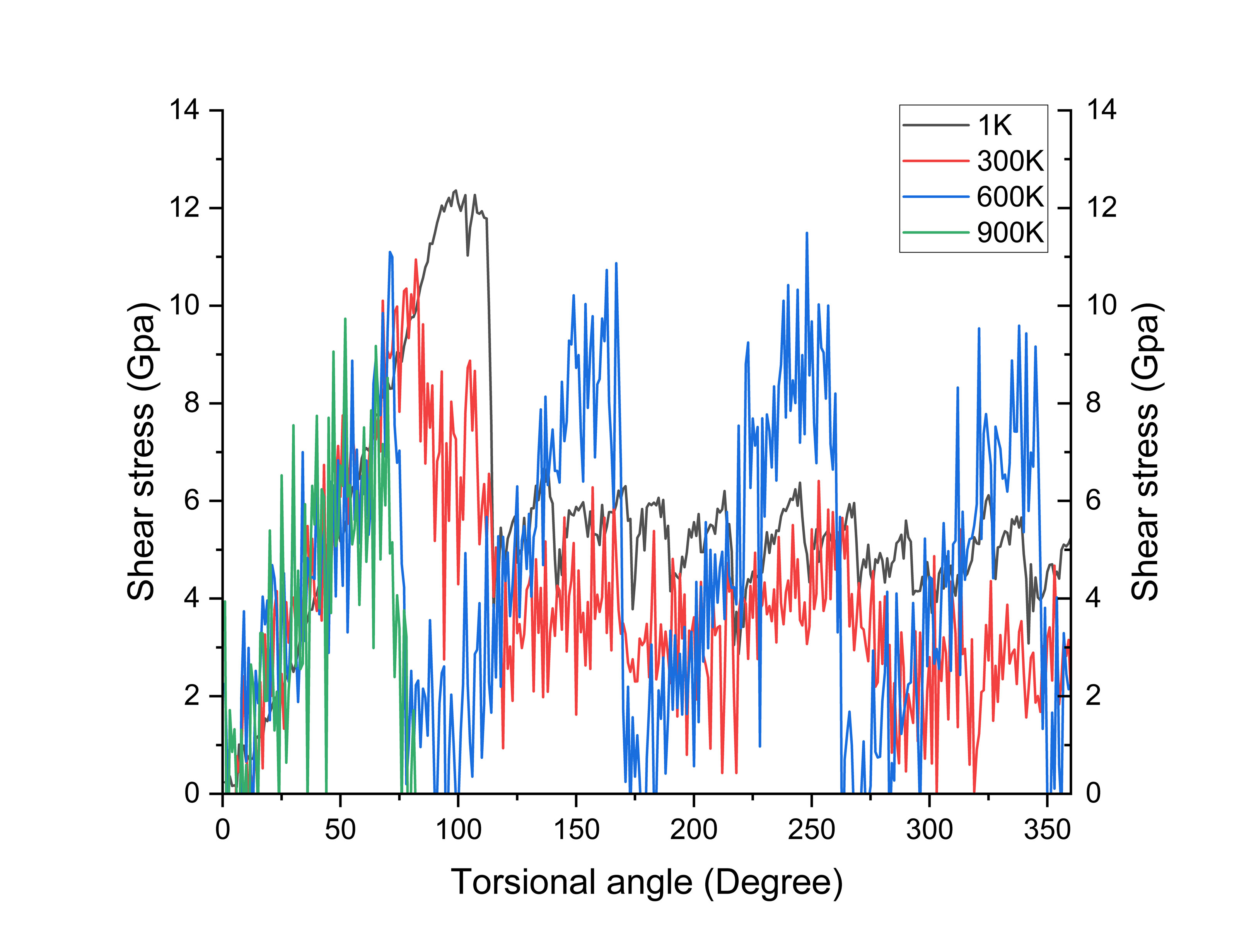}
	\caption{Shear stress vs. torsional angle for a 10\(a\) nanowire with 1\% carbon concentration. Each curve corresponds to a specific temperature (1 K, 300 K, 600 K, and 900 K}
	\label{tmp02}
\end{figure}

\begin{figure}[H]
	\centering
	\includegraphics[width=0.5\textwidth]{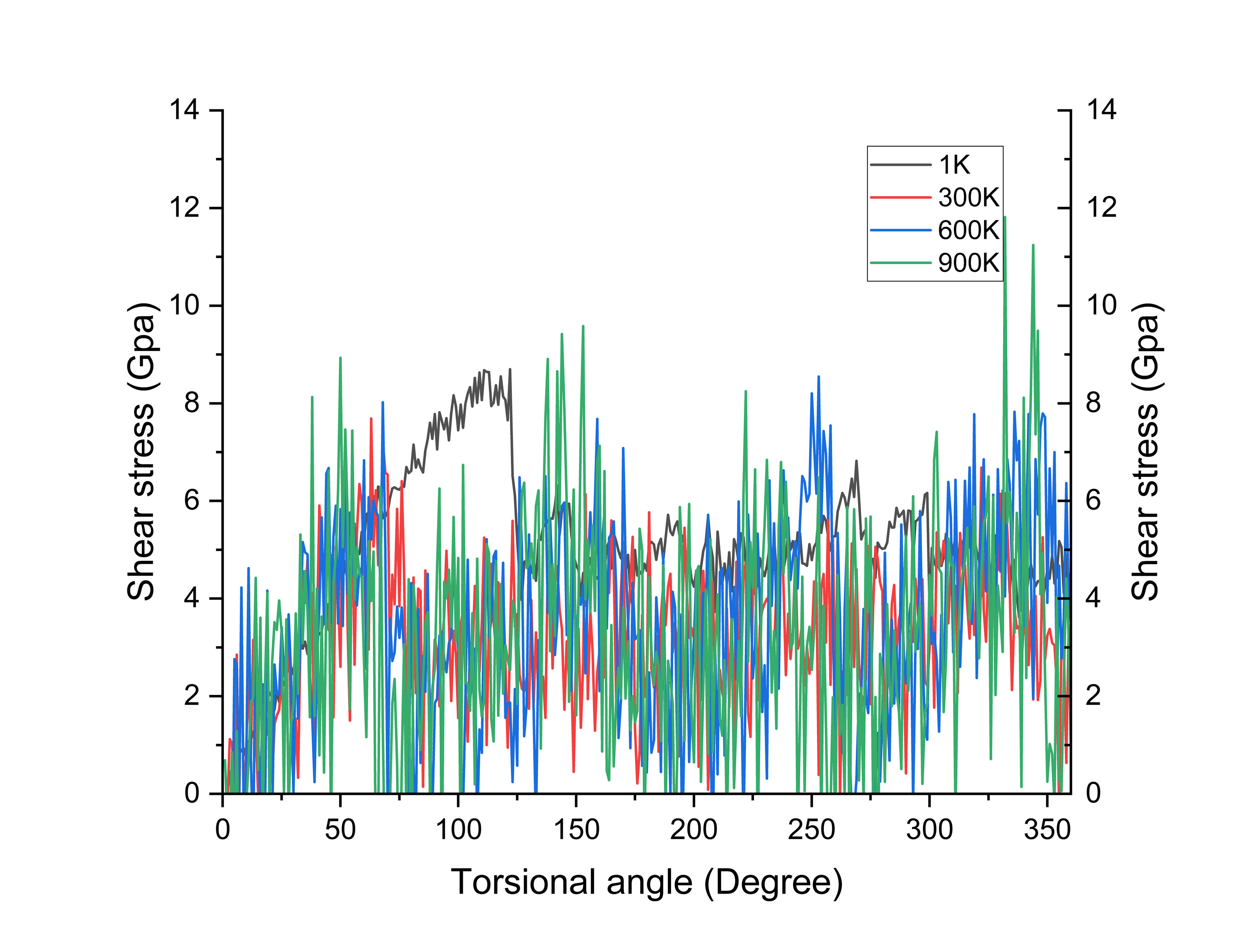}
	\caption{Shear stress vs. torsional angle for a 10\(a\) nanowire with 5\% carbon concentration. Each curve corresponds to a specific temperature (1 K, 300 K, 600 K, and 900 K}
	\label{tmp03}
\end{figure}

\begin{figure}[H]
	\centering
	\includegraphics[width=0.5\textwidth]{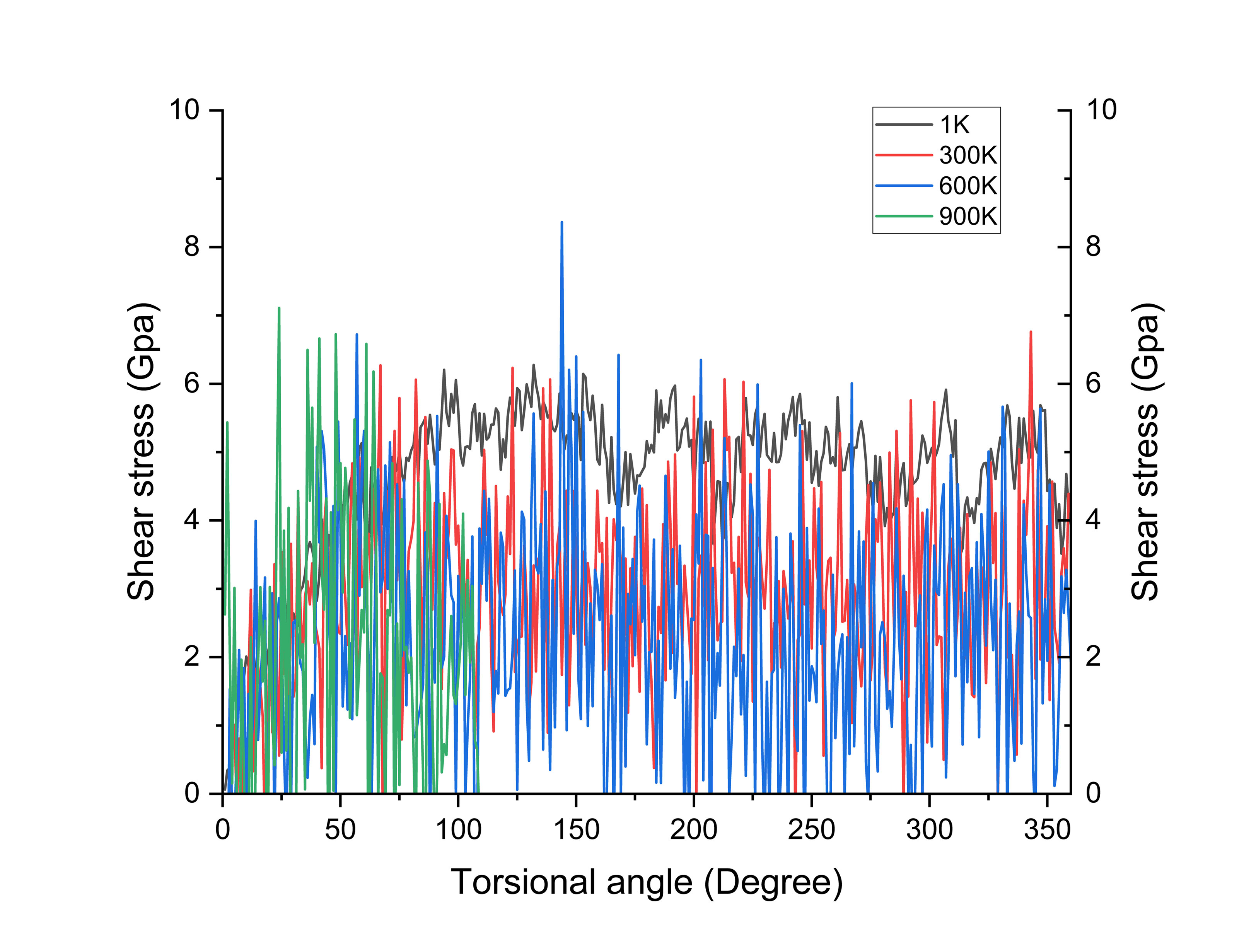}
	\caption{Shear stress vs. torsional angle for a 10\(a\) nanowire with 10\% carbon concentration. Each curve corresponds to a specific temperature (1 K, 300 K, 600 K, and 900 K}
	\label{tmp04}
\end{figure}

Figure \ref{tmp01} shows the shear stress versus torsional angle for a nanowire with 0\% carbon concentration. At 1 K, the nanowire exhibits a high critical angle of approximately 140°, and the transition from the elastic to plastic region is clearly identifiable. However, as the temperature increases, the critical angle reduces gradually, and minor stress fluctuations appear due to increased thermal activity. In Figure \ref{tmp02}, the nanowire with 1\% carbon concentration shows a similar trend, with the critical angle at 1 K around 100°, followed by a decrease as temperature rises. The increase in fluctuations at higher temperatures also becomes more prominent. Figure \ref{tmp03} illustrates the torsional response of a nanowire with 5\% carbon. While the critical angle at 1 K remains distinguishable at approximately 114°, the stress-strain curves become unstable with increasing temperature, making it difficult to clearly identify the critical points at elevated conditions. Figure \ref{tmp04}, which shows the 10\% carbon nanowire, reveals a highly unstable torsional behavior across all temperatures, including 1 K, with significant fluctuations and indistinct transition points between deformation stages.

\begin{figure}[H]
	\centering
	\includegraphics[width=0.4\textwidth]{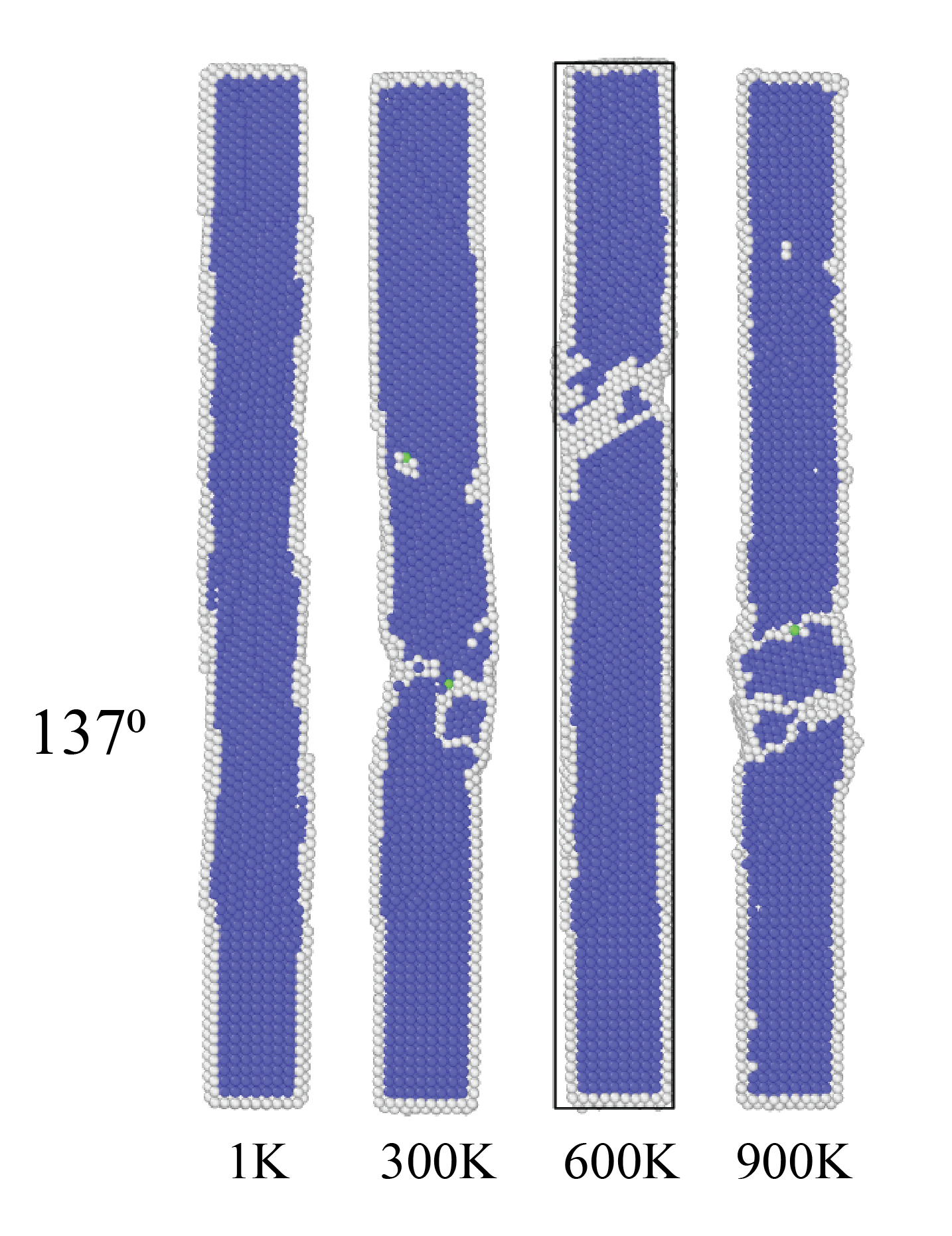}
	\caption{Dislocation structures of the model at a 137° torsional angle under varying temperatures.}
	\label{Cross tem diff}
\end{figure}

Additionally, as carbon concentration increases, the structure exhibits more dislocations, further weakening the structural integrity of the nanowire under torsional loading. Figure \ref{Cross tem diff} presents the atomic dislocation configurations for nanowires twisted to 137° at different temperatures. At 0 K, the nanowire retains its structural order with minimal dislocation activity. In contrast, higher temperatures significantly enhance dislocation density, especially at 900 K, where the structure shows extensive atomic disorder and deformation. These results demonstrate that thermal motion plays a crucial role in destabilizing the atomic structure during torsion at elevated temperatures.

\newpage
\

\textbf{Effect of Cross-Sectional Size on Torsion}\\

\begin{figure}[H]
	\centering
	\includegraphics[width=0.5\textwidth]{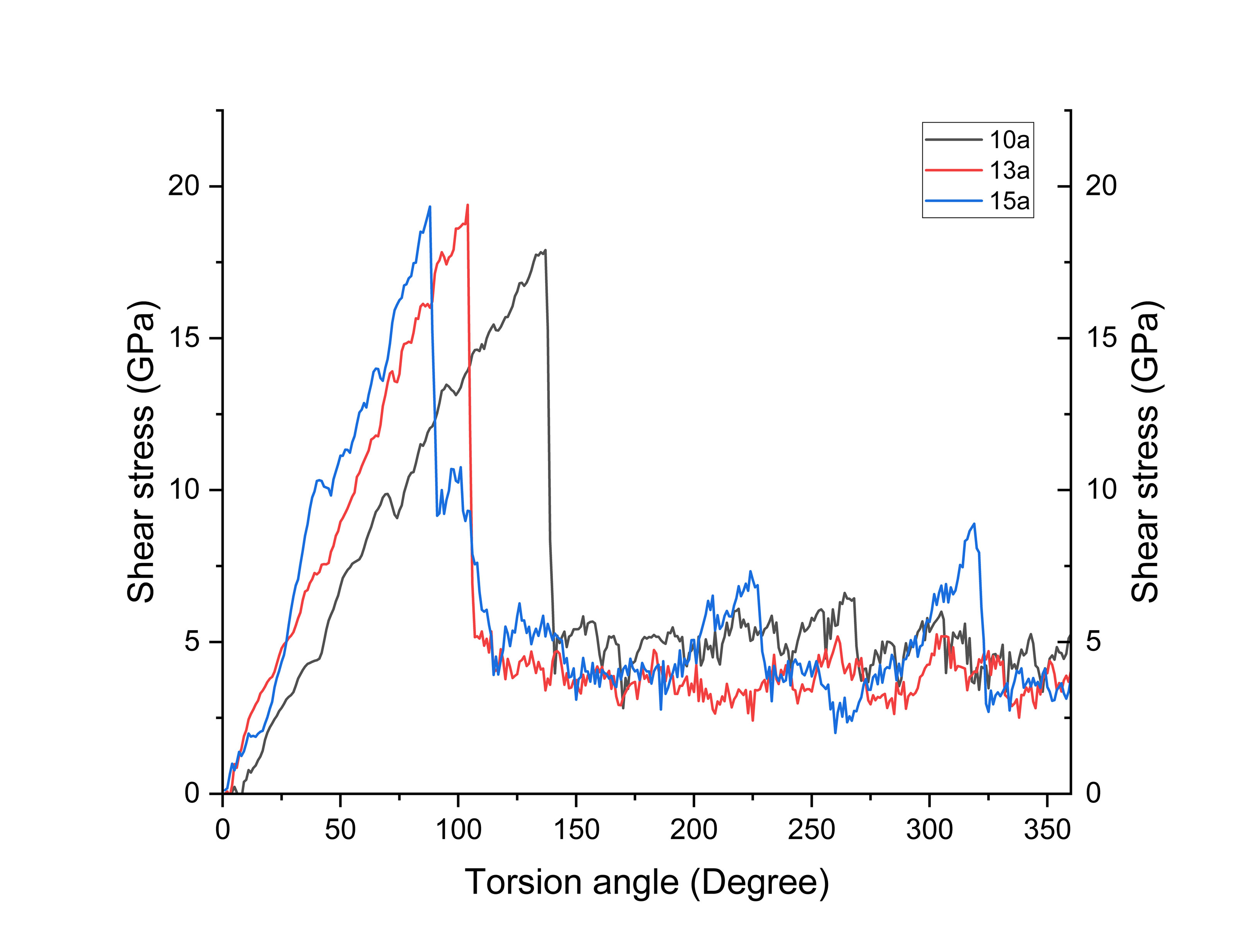}
	\caption{Shear stress vs. torsional angle at 1 K for nanowires with 0\% carbon concentration. Each curve corresponds to a different cross-sectional radius (10\(a\),13\(a\),and 15\(a\))}
	\label{cross1}
\end{figure}

\begin{figure}[H]
	\centering
	\includegraphics[width=0.5\textwidth]{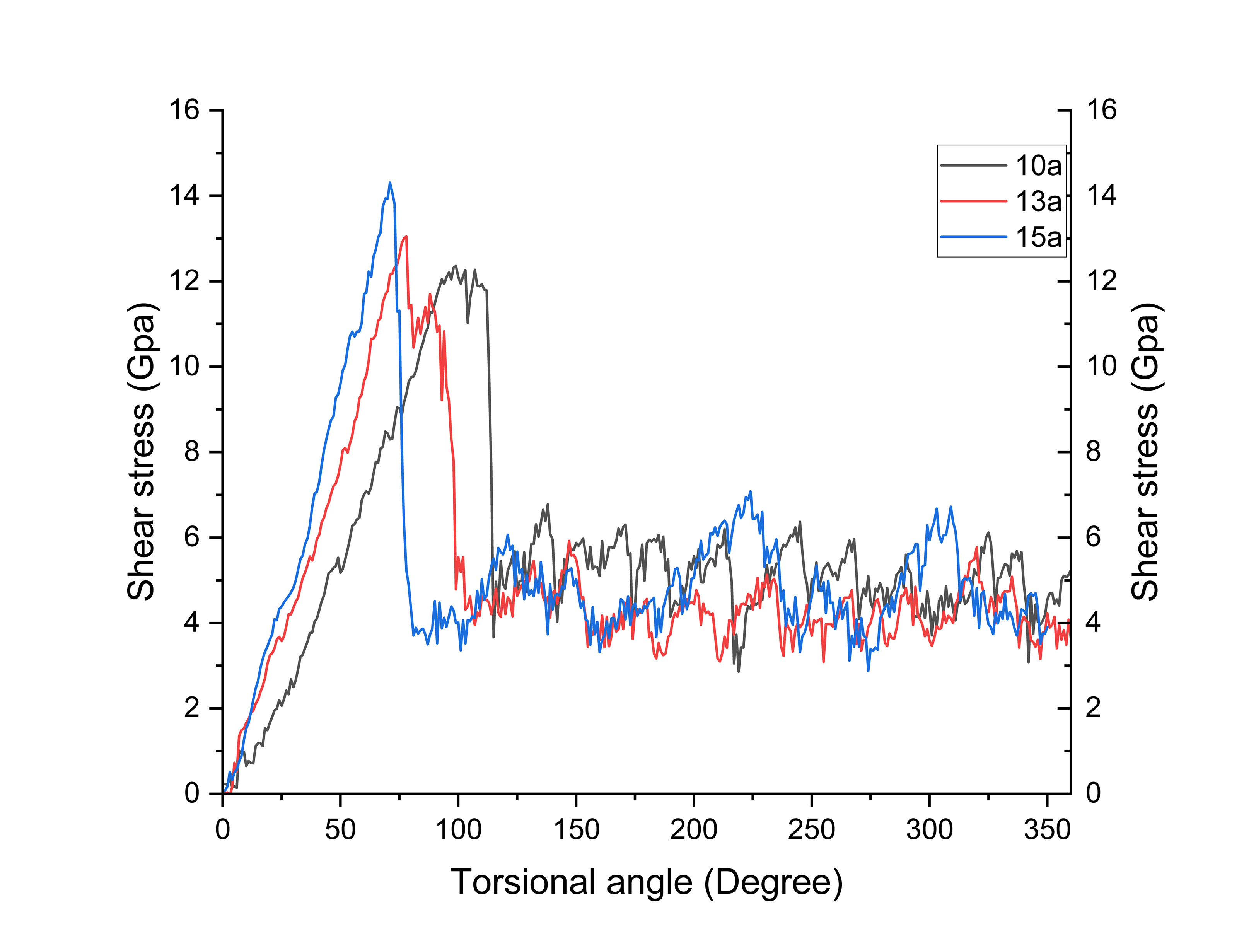}
	\caption{Shear stress vs. torsional angle at 1 K for nanowires with 1\% carbon concentration. Each curve corresponds to a different cross-sectional radius (10\(a\),13\(a\),and 15\(a\))}
	\label{cross2}
\end{figure}

\begin{figure}[H]
	\centering
	\includegraphics[width=0.5\textwidth]{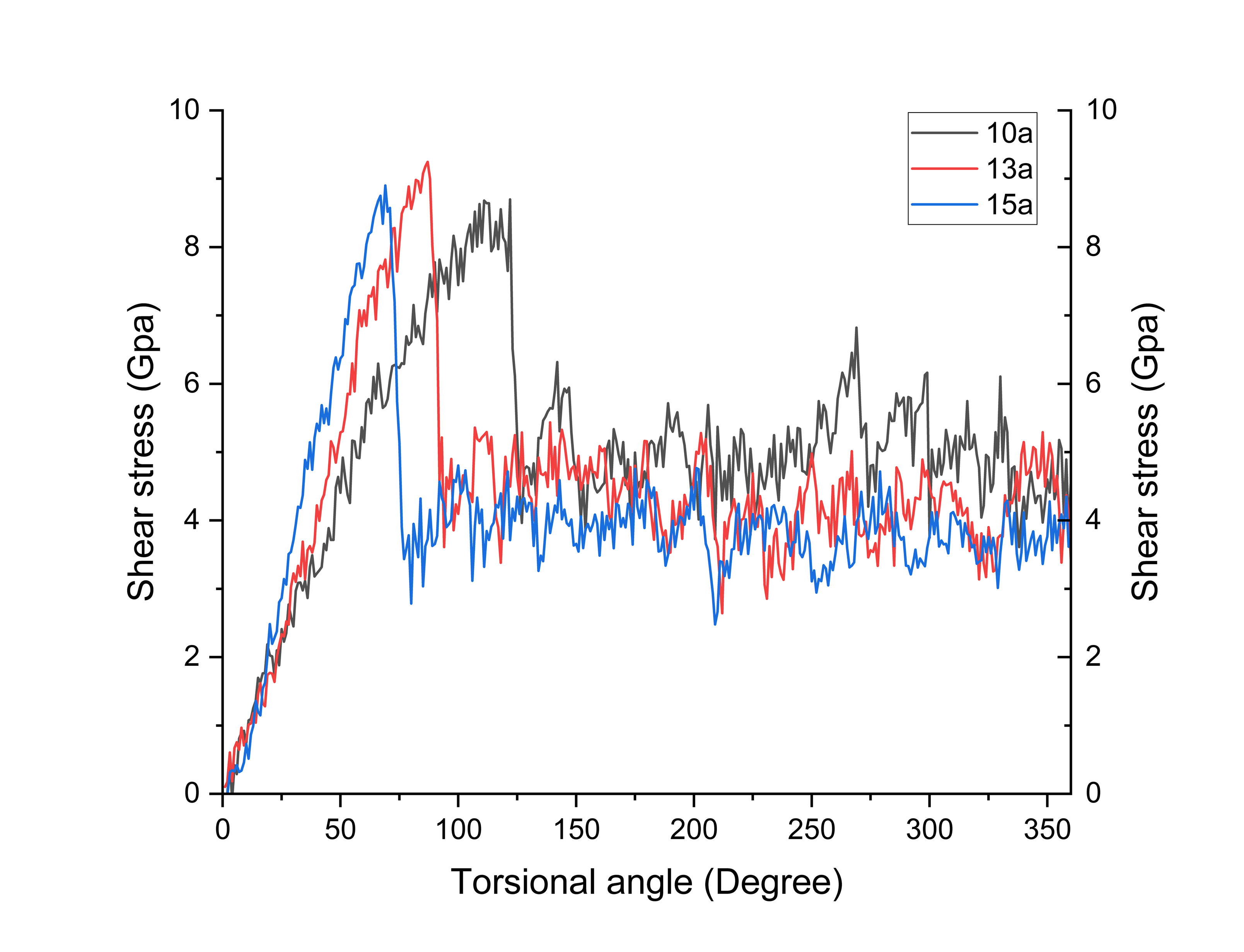}
	\caption{Shear stress vs. torsional angle at 1 K for nanowires with 5\% carbon concentration. Each curve corresponds to a different cross-sectional radius (10\(a\),13\(a\),and 15\(a\))}
	\label{cross3}
\end{figure}

\begin{figure}[H]
	\centering
	\includegraphics[width=0.5\textwidth]{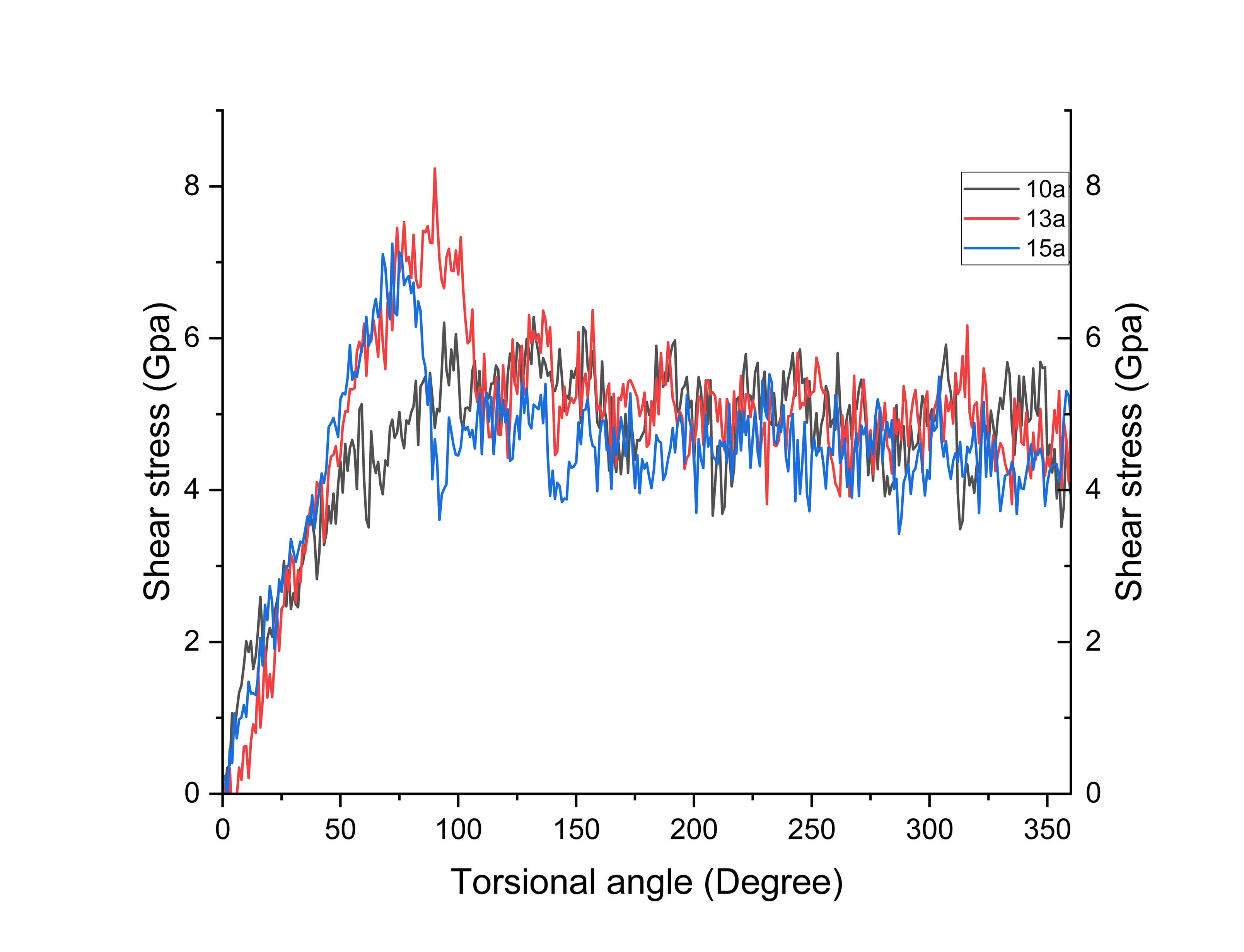}
	\caption{Shear stress vs. torsional angle at 1 K for nanowires with 10\% carbon concentration. Each curve corresponds to a different cross-sectional radius (10\(a\),13\(a\),and 15\(a\))}
	\label{cross4}
\end{figure}

The influence of cross-sectional size on the torsional response of Fe–C nanowires was analyzed at a constant temperature of 1 K, considering different carbon concentrations. Figure \ref{cross1} shows the shear stress versus torsional angle for nanowires with 0\% carbon concentration. The results indicate that as the cross-sectional diameter increases, the critical torsion angle — representing the transition from elastic to plastic deformation — decreases. However, the shear stress required to reach the critical angle increases significantly in larger nanowires compared to smaller ones. This trend is also observed in Figure \ref{cross2}, which presents the torsional behavior of nanowires with 1\% carbon concentration. Larger nanowires require higher shear stress to reach lower critical angles, following a similar pattern to the undoped case. Figure \ref{cross3} illustrates the behavior at 5\% carbon concentration, where the critical angle is again lower in nanowires with larger cross-sections. However, in this case, the shear stress required to reach the critical point is nearly identical across all cross-sectional sizes, suggesting a balance between geometry and impurity effects. In contrast, Figure \ref{cross4} presents the torsional response of nanowires with 10\% carbon content. The results show unstable and highly fluctuating stress responses, making it difficult to determine clear critical angles. Notably, the 13a cross-section nanowire achieves the highest recorded stress value, but due to the irregular fluctuations at high carbon content, the data lacks reliable consistency. These results reveal a general trend in which increased cross-sectional size correlates with maximum shear stress, despite a reduced critical torsion angle. This behavior is attributed to the larger outer surface area of thicker nanowires, which requires greater torque to initiate deformation.

\begin{figure}[H]
	\centering
	\includegraphics[width=0.5\textwidth]{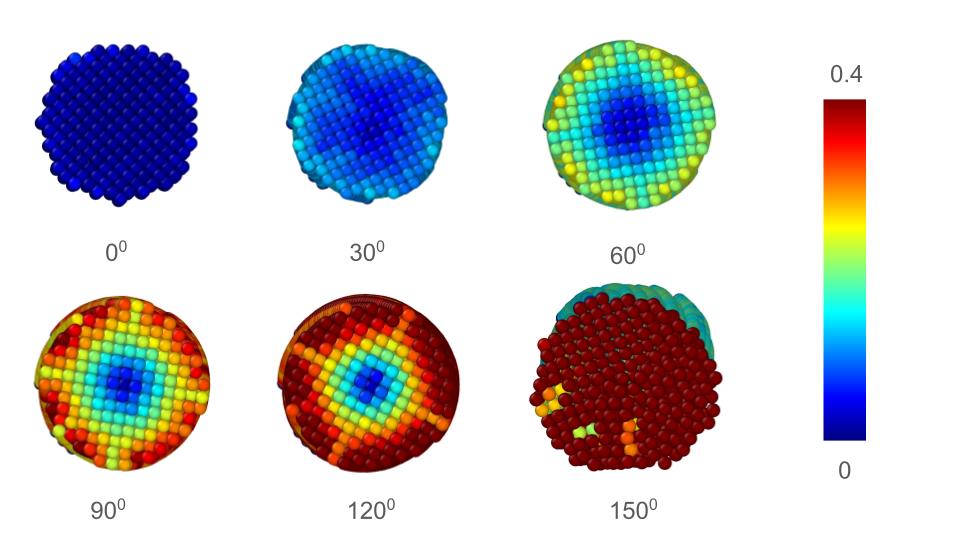}
	\caption{Strain distribution across the cross-section of a 10 a nanowire.}
	\label{fig: strain Dis}
\end{figure}

At the atomic scale, torsional deformation begins at the outermost atomic layers and propagates inward. As shown in Figure \ref{fig: strain Dis}, the strain distribution across the cross-section confirms that the outer layers experience the highest strain under torsional loading, making them the primary sites for bond breakage and the initiation of plastic deformation. Consequently, nanowires with larger cross-sections possess more atoms in these high-strain regions, resulting in increased resistance to torsional deformation. However, this geometric effect is influenced by material-specific factors. Higher carbon concentrations introduce lattice distortions that weaken the structural integrity, lowering the stress required for dislocation nucleation and reducing mechanical stability. Similarly, increased temperature reduces atomic bonding strength, further facilitating deformation. Overall, these findings highlight the complex interplay between geometrical factors, such as cross-sectional size, and material characteristics, including carbon concentration and temperature. Together, these parameters govern the torsional mechanical performance of Fe–C nanowires and offer valuable guidance for the design and optimization of nanoscale structures subjected to torsional loads.

\section*{Conclusion}

Molecular dynamics simulations were conducted to systematically investigate the torsional mechanical behavior of ferrous (Fe) and carbon-doped ferrous (Fe–C) nanowires under varying carbon concentrations, temperatures, and cross-sectional sizes. The analysis focused on the shear stress–torsion angle response, critical torsion angles, and dislocation evolution during deformation. The results reveal that the mechanical response of the nanowires is highly sensitive to these parameters, with clear trends observed in torsional strength and deformation stability.

The simulations demonstrate that increasing carbon concentration weakens the nanowire’s ability to sustain torsional loads. As the concentration rises, the maximum shear stress at the critical angle decreases, and the structural integrity becomes increasingly unstable. At higher doping levels, the distinction between elastic and plastic deformation stages becomes unclear, indicating severe lattice mismatch and early onset of dislocation nucleation.

Temperature also plays a critical role in influencing the mechanical properties. At lower temperatures, nanowires exhibit higher torsional strength and larger critical angles. As temperature increases, thermal vibrations enhance atomic mobility, reducing the shear strength and promoting earlier plastic deformation. This thermal effect significantly impacts the structural stability of the nanowires under torsional loading, particularly at high carbon concentrations.

In contrast, increasing the cross-sectional size of the nanowires enhances their torsional resistance. Larger cross-sections require greater shear stress to initiate deformation due to a greater number of atomic layers at the outer surface, where torsion-induced strain is most pronounced. However, this size-dependent strengthening effect is moderated by material parameters such as carbon content and temperature.

Overall, the combined effects of doping, geometry, and temperature define the torsional performance of Fe–C nanowires. The findings provide useful insight into the mechanical behavior of nanoscale materials and contribute to the understanding required for the design and optimization of nanowire-based components in nanoelectromechanical systems and advanced functional materials. These results may serve as a reference for future studies exploring the torsional mechanics of complex nanostructures.

\bibliographystyle{plainnat} 
\bibliography{Reference.bib}






\end{document}